\documentclass[twocolumn,superscriptaddress,amsmath,amssymb,showpacs,pre]{revtex4-1}
\usepackage{graphicx}
\usepackage{color}
\renewcommand{\phi}{\varphi}

\begin{document}

\title{Competitive B2 and B33 Nucleation during Solidification of Ni$_{\text{50}}$Zr$_{\text{50}}$ Alloy: \\
Molecular Dynamics Simulation and Classical Nucleation Theory}

\author{Yang Sun}
	\affiliation{Ames Laboratory, US Department of Energy, Ames, Iowa 50011, USA}
\author{Feng Zhang}
	\affiliation{Ames Laboratory, US Department of Energy, Ames, Iowa 50011, USA}
\author{Huajing Song}
	\affiliation{Ames Laboratory, US Department of Energy, Ames, Iowa 50011, USA}
\author{Mikhail I. Mendelev}
	\affiliation{Ames Laboratory, US Department of Energy, Ames, Iowa 50011, USA}
\author{Cai-Zhuang Wang}
	\affiliation{Ames Laboratory, US Department of Energy, Ames, Iowa 50011, USA}
	\affiliation{Department of Physics, Iowa State University, Ames, Iowa 50011, USA}
\author{Kai-Ming Ho}
	\affiliation{Ames Laboratory, US Department of Energy, Ames, Iowa 50011, USA}
	\affiliation{Department of Physics, Iowa State University, Ames, Iowa 50011, USA}

\date{Feb. 23, 2019}

\begin{abstract}
We investigated the homogenous nucleation of the stoichiometric B2 and B33 phases in the Ni$_{\text{50}}$Zr$_{\text{50}}$  alloy using the persistent embryo method and the classical nucleation theory. The two phases become very close competitors at large supercoolings, which is consistent with the experimental observations. In the case of the B2 phase, the linear temperature dependence of the solid-liquid interface (SLI) free energy extrapolated to the melting temperature leads to the same value as the one obtained from the capillarity fluctuation method (CFM). In the case of the B33 phases, the SLI free energy is also a linear function of temperature at large supercoolings but the extrapolation to the melting temperature leads to a value which is considerably different from the CFM value. This is consistent with the large anisotropy of the SLI properties of the B33 phase nearby the melting temperature observed in the simulation of the nominally flat interface migration.

\end{abstract}

\maketitle

\section{Introduction}

The nucleation rate is one of the key factors necessary to understand and predict the phase competition and transformation pathways \cite{R1,R2}. The classical nucleation theory (CNT) is widely used to describe the nucleation process. The simplest scenario within this theory assumes formation of a crystal nucleus as result of thermal fluctuations in the liquid phase. Once this nucleus reaches a critical size it will grow until all liquid solidifies or it meets another growing crystal. The situation becomes more complex when different crystal phases can nucleate within the liquid phase. Obviously, only one of these crystal phases is stable and all others are metastable. Yet, this does not mean that it is the stable phase which will nucleate first. For example, it has been shown that the body-centered cubic (bcc) phase can nucleate prior to the face-centered cubic (fcc) phase in the systems where the fcc phase is the most stable phase \cite{R3,R4,R5,R6,R7,R8,R9,R10}. The CNT can explain this phenomenon assuming that the bcc-liquid interface free energy is smaller than the fcc-liquid interface free energy in the same system \cite{R11,R12}. 

Making a specific prediction based on the CNT is a much more challenging problem because any specific CNT prediction depends on the input information which is usually not available. For example, if a metastable phase exists only for a relatively short time, it is very difficult to get even the bulk thermodynamic data for this phase from experiments. The determination of the solid-liquid interface (SLI) free energy for such a phase from experiment is almost impossible. The anisotropy of SLI free energy and mobility brings further complication because one should take into account the nucleus shape which may not be spherical. Without this information the CNT cannot be used to predict the outcome of the phase competition.

Some of the problems mentioned above can be solved by coupling CNT with molecular dynamics (MD) simulation. Since such an MD simulation requires utilizing semi-empirical potentials, the outcome may heavily depend on the quality of these potentials. This issue has been discussed in the literature (e.g., see \cite{R13,R14}) and we will not focus on it in the present paper. However, even if there is a perfect semi-empirical potential (which is never the case) there are still many obstacles for the application of the CNT. First of all, while the difference between the bulk free energies of the solid and liquid phases can be relatively easily and reliably obtained from the MD simulation \cite{R15,R16,R17}, the measurement of the SLI free energy, $\gamma$, is a long-standing problem. A well-established approach to compute $\gamma$ is the capillary fluctuation method (CFM) \cite{R18} where SLI stiffness is determined at the melting temperature based on the capillary wave theory  \cite{R19,R20}. Laird and co-workers further extended the CFM along the pressure-temperature coexistence curve using the “Gibbs-Cahn integration” method \cite{R21}. However, the application of CFM at the ambient pressure well below the melting temperature is still problematic. 

The prediction of nucleus shape is an additional problem. Although one can get the SLI free energy as a function of interface orientation from the CFM calculations, we are not aware of actual comprehensive MD studies where all of these parameters were obtained and utilized within the CNT. Moreover, it is still not clear whether the SLI data obtained for nominally flat interfaces can be used in the nucleation studies especially when the nucleus size is small and special corrections are needed (e.g., see  \cite{R22,R23,R24}). 

An alternative approach is to obtain the nucleation rate directly from the MD simulation. Unfortunately, it has been done only for pure metals \cite{R25} and for some binary alloys (e.g., see \cite{R26}) simply because it usually takes too long (more than 1 $\mu s$) to nucleate in a simulation cell containing hundreds of thousands of atoms. This problem can now be overcome using the persistent embryo method (PEM) developed in \cite{R27} . In this method a subcritical crystal nucleus is inserted in the liquid and kept from melting by springs which are gradually removed once the nucleus reaches a threshold size which is still smaller than the critical size. With this method, even the nucleation in glass forming metallic alloys can be studied\cite{R27,R35, R52}. The outcome of the MD simulation in the PEM is the critical nucleus size, $N^{*}$, and shape factor, $s^{*}$ (which will be defined in Section 2). Along with the bulk thermodynamic driving force, $\Delta\mu$ this information is sufficient to get the nucleation rates using the CNT. 
	
The information obtained using the PEM can be also utilized to extract the temperature dependence of the SLI free energy  \cite{R28}. This approach relies on two very strong approximations. First, contrary to the CFM, it ignores the orientation dependence of the SLI free energy and provides an estimate for the “average” value of SLI free energy. Thus, the question about how the anisotropy of SLI free energy changes with the temperature remains open. Second, as was discussed in \cite{R28}, the obtained value includes the effects associated with larger interface curvature of the critical nucleus which is usually rather small. Yet, at the present moment, this is the simplest approach to get this dependence from the atomistic simulation. 

The temperature dependences of the SLI free energy of pure Al and Ni were obtained in \cite{R28} using the PEM. In both cases, the temperature dependences were pretty linear. The extrapolation of these linear dependences to the melting temperatures showed a good agreement with the average SLI free energy obtained using the CFM which demonstrate a reliability of the obtained data. However, it is not obvious that similar linear dependences will hold for more complicated multi-component phases.

The PEM is an extension of the seeding method and for the sake of completeness we now briefly review other seeding method studies where the temperature dependence of the SLI free energy was obtained. Bai and Li used the seeding method to obtain $\gamma(T)$ for the Lennard-Jones system \cite{R29}. They also found that the temperature dependence is linear. Sanz et al. further extended the method to determine $\gamma(T)$ in a LJ system, water and NaCl \cite{R30}. They obtained linear temperature dependences for the LJ system and water while the data obtained for NaCl were too scattered to make any conclusion.
A clear observation from the nucleation studies mentioned above is that the temperature dependence of the SLI free energy can be very important in the prediction of the nucleation rate such that the values obtained using the CFM at the melting temperature cannot be used for this purpose. On the other hand, the CFM is a well-developed technique such that it would be very fruitful to find a way to utilize the CFM data in the CNT. In order to do it an empirical temperature dependence of the SLI free energy can be used. Trying to find an empirical rule to predict the SLI free energy for different systems, Turnbull proposed to relate it with the latent heat, $\Delta H$, and atomic density, $\rho_{c}$, as \cite{R31}
\begin{equation}
\gamma=C_T \rho_{c}^{\frac{2}{3}} \Delta H
\label{eos}.
\end{equation} 
where $C_T$ is a constant which depends on the type of the crystal lattice. Using this relation to evaluate the temperature dependence of the SLI free energy seems to be a straightforward approach \cite{R25,R29}. However, the comparisons between the temperature dependences of the SLI free energies obtained from the seeding method or the PEM with the predictions based on the Turnbull relation do not show a good agreement between them \cite{R28,R30,R32}. 

The goal of the present study was to apply the PEM and the CNT to describe the phase competition during the solidification of the Ni$_{\text{50}}$Zr$_{\text{50}}$ alloy. The solidification of this alloy was recently studied using the electrostatic levitation technique \cite{R33}. The experiment showed that while B33 is the most stable crystal phase from $T=0$ to the melting temperature, a metastable B2 phase was found to nucleate first. This phenomenon was explained in  \cite{R33} by the fact that the B2-liquid interface free energy is much smaller than the B33-liquid interface free energy. However, this conclusion was made based on the CFM calculations performed at the melting temperatures and the Turnbull relation was used to extrapolate the obtained values down to the large supercoolings. At the same time, the data obtained for the SLI velocity in \cite{R34} indicate that the temperature dependences of the SLI properties in this alloy can be very non-trivial (see Fig. 5 in \cite{R34}): the B33 SLI normal to the [010] direction almost did not move during MD simulation near by the melting temperature while the SLIs with other orientations were mobile at the same temperatures. On the other hand, the [010] SLI velocity was found to be comparable with velocities for other orientations at large undercoolings. While no correlation between the temperature dependences of the SLI free energy and velocity has been established so far it is reasonable to expect that the temperature dependence of the SLI free energy in the Ni$_{\text{50}}$Zr$_{\text{50}}$ alloy should be also non-trivial.

The rest of the paper is organized as follows. In the next section, we will describe the classical nucleation theory, the persistent embryo method, the order parameters, and the simulation details. In Section 3, we will present the obtained temperature dependences of the critical nucleus sizes and the SLI free energies for B2 and B33 phase. In Section 4, we will use these data to compute the nucleation rate in a wide temperature range for both phases. Finally, in Section 5 and 6, we will discuss and summarize the obtained results.

\section{Persistent-embryo method}
According to the CNT \cite{R1}, a homogeneous nucleation involves a formation of a critical nucleus in an undercooled liquid. The formation of such a nucleus is governed by the thermodynamic driving force and the energy penalty associated with creating a solid-liquid interface between the nucleus and the liquid. The excess free energy to form the nucleus consisting of N atoms is 
\begin{equation}
\Delta G = N \Delta \mu + A \gamma
\label{eos},
\end{equation}
where $\Delta \mu (<0)$ is the chemical potential difference between the bulk solid and liquid phases, $\gamma$ is the solid-liquid interfacial free energy, and A is the interface area which can be evaluated as $ A = s ( \frac{N}{\rho_c} ) ^ \frac{2}{3} $, where $\rho_c$ is the crystal density and $s$ is a shape factor \cite{R28}. The competition between the bulk and interface terms leads to a nucleation barrier, which can be determined from the condition that $\frac{\partial \Delta G (N^{*})}{\partial N} = 0$:
\begin{equation}
\Delta G^{*} = \frac{4 s^{*3} \gamma^{3} }{ 27 \vert \Delta \mu \vert ^{2} \rho_{c}^{2}} 
\label{eos}.
\end{equation}

Typically, the CNT assumes the spherical shape ($s_{CNT}\equiv \sqrt[3]{36 \pi}$) of the nucleus to relate $\Delta G^{*}$ with $\gamma$ and $\Delta \mu$. This assumption can be soften by supposing that the averaged shape of the sub-critical nucleus does not change at the critical size. This supposition leads to the following expressions for the nucleation barrier and the SLI free energy \cite{R27}:
\begin{equation}
\Delta G^{*} = \frac{1}{2} \vert \Delta \mu \vert N^{*}
\label{eos}.
\end{equation}
and
\begin{equation}
\gamma=\frac{3}{2s^\ast}\Delta \mu \rho_{c}^{\frac{2}{3}} N^{*\frac{1}{3}}
\label{eos}.
\end{equation} 

\begin{figure}
\includegraphics[width=0.48\textwidth]{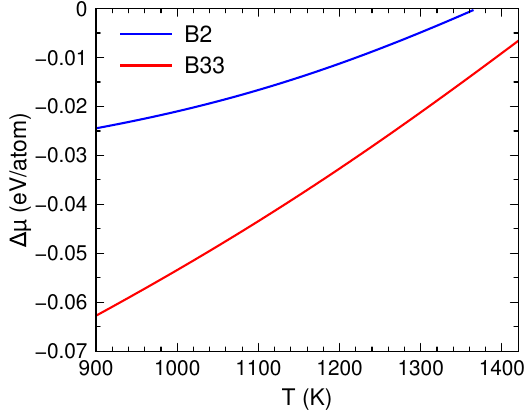}
\caption{\label{fig:fig1} The bulk driving force for the solidification in the Ni$_{\text{50}}$Zr$_{\text{50}}$ alloy. }
\end{figure}

Equation (5) shows that four quantities ($\rho_c$, $\Delta \mu$, $N^{*}$, and $s^{*}$) are needed to obtain from MD simulations to calculate the interfacial free energy $\gamma$ at a given temperature. The crystal density, $\rho_c$,  and the melting temperatures for the B2 and B33 phases in the Ni$_{\text{50}}$Zr$_{\text{50}}$ alloy have been obtained in \cite{R34}. The methods to determine the liquid and crystal densities and the melting temperature were described in details in \cite{R53} and in \cite{R54}, respectively. The melting temperatures for B2 and B33 phases were found to be $1369 K$ and $1473 K$, respectively. The bulk driving force $\Delta \mu$ at a target temperature $T_t$ was calculated using the Gibbs-Helmholtz equation: 
\begin{equation}
\Delta \mu (T_t)=T_t  \int_{T_m}^{T_t} \frac{\Delta H(T)}{T^2}dT
\label{eos},
\end{equation} 
where $T_m$ is the melting temperature and $\Delta H(T)$ is the latent heat which is obtained directly by the MD simulations. The value of $\Delta \mu$ are shown as a function of temperature in  Fig.~\ref{fig:fig1} for both B2 and B33 phases. Examination of this figure shows that the employed potential does provide in agreement with experiment that the B33 phase is always more stable than the B2 phase.

\begin{figure}
\includegraphics[width=0.48\textwidth]{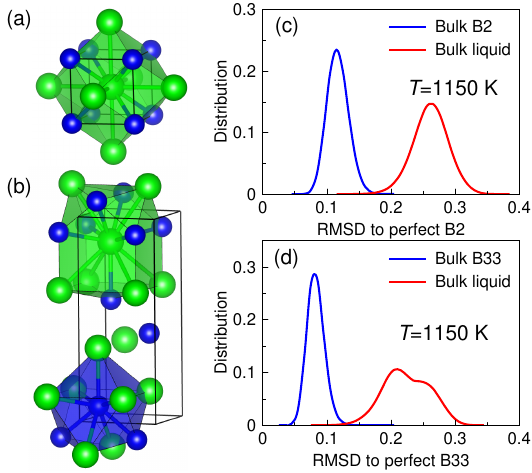}
\caption{\label{fig:fig2} Using the cluster alignment method to differentiate the B2 and B33 phases from the liquid. (a) and (b) The main polyhedra in the B2 and B33 crystals, respectively. The green balls represent Zr and the blue balls represent Cu. (c) and (d) The RMSD for the separate models of bulk phases.}
\end{figure}

The critical nucleus size, $N^{*}$, and the shape factor, $s^{*}$, were obtained in the present study using the PEM. The PEM utilizes the main CNT concept that a homogeneous nucleation happens via the formation of the critical nucleus in an undercooled liquid. The nucleation process can be efficiently sampled in the PEM  by preventing a small crystal embryo consisting of $N_0$ atoms ($N_0 \ll N^*$) from melting using external spring forces \cite{R27, R35,R36}. Very long ineffective simulations, in which the system is very far away from forming the critical nucleus, are avoided in this method. As the embryo grows, the harmonic potential is gradually weakened and is completely removed when the cluster size reaches a sub-critical threshold, $N_{sc}$  ($N_0<N_{sc}<N^{*}$). If the nucleus melts below $N_{sc}$ the harmonic potential is gradually enforced preventing the complete melting of the embryo. Since the harmonic potential is applied only to $N_0$ atoms of the original embryo and it is removed well before the nucleus reaches the critical size this harmonic potential does not affect the critical nucleus shape and its interface with the liquid phase. Thus, the system is unbiased at the critical point and reliable information about the critical nucleus can be obtained. Our method does not require fine tuning of $N_0$ and $N_{sc}$ \cite{R27} provided that the choice of these parameters is reasonable. When $N_0$ and $N_{sc}$ are too large, the crystallization takes place so fast that one can barely detect any clear critical plateau from simulation trajectories. When $N_0$ and $N_{sc}$ are too small, no nucleation event or any critical plateau can be observed within the simulation time (see details in the supplementary of Ref. \cite{R27}). A reasonable choice of $N_0$ and $N_{sc}$ is achieved by a few trial-and-error cycles until the critical fluctuating plateau can be observed within typical MD simulation time. 
When the nucleus reaches the critical size, it has equal chances to melt or to further grow causing fluctuations around $N^*$. As a result, the $N(t)$ curve tends to display a plateau during the critical fluctuations, giving a unique signal to detect the appearance of the critical nucleus. Sometimes, several values of $N^*$ can be obtained from the same simulation if the critical nucleus melt rather than grow. An additional statistic can be also obtained by changing the initial atomic velocities. At the same time, sufficient statistics on the critical nucleus shape and the shape factor, $s^*$, can be obtained from the same simulations. \cite{R27,R28} 
	To identify the solid-like atoms from liquid on the fly, we employed the efficient kinetic parameter \cite{R37} which is based on the analysis of the atomic displacements within a specified time interval. The atoms which show vibrational motion are recognized as solid, while the ones showing diffusional motion are recognized as liquid. The clustering analysis \cite{R38}, which uses the crystalline bond length as the cutoff distance to choose neighbor solid atoms, was employed to measure the size of the nucleus. Since the nucleus size can be rather sensitive to the choice of order parameters \cite{R39}, in addition to the kinetic parameter, the cluster-alignment (CA) method \cite{R40} was employed to validate the nucleus size based on the solid structures. Being a structural order parameter, the CA method can well differentiate complex crystal structures by computing the minimal root-mean-square deviation (RMSD) between the atomic cluster and the perfect crystal motifs \cite{R37,R41}. The bcc polyhedron was chosen as the crystal template (see Fig.~\ref{fig:fig2}a) for the B2 phase. In the case of the B33 phase we took into account that the Ni and Zr atoms have different local environments. Therefore, we chose the polyhedra formed by the closest neighbor atoms near the Ni and Zr atoms as shown in Fig.~\ref{fig:fig2}(b). Figures~\ref{fig:fig2}(c) and (d) show that the RMSDs between the B2 and liquid phases, as well as the B33 and liquid phases, are well separated from each other. It was found the critical nucleus size determined from the cluster alignment is rather consistent with the one determined from the kinetic order parameter.
	
All MD simulations in the present study were performed using the GPU-accelerated LAMMPS code \cite{R42,R43,R44}. The Ni-Zr Finnis-Sinclair potential \cite{R45} developed in \cite{R34} was employed. All MD simulations were performed using the NPT ensemble with the Nose-Hoover thermostat and Parinello-Rahman barostat. The time step of the MD simulation was $1.0 fs$. The damping time in the Nose-Hoover thermostat was set as $\tau =0.1 ps$, which is frequent enough for the heat dissipation during the crystallization \cite{R28}. The simulation cell contained 31,250 atoms which is at least 20 times larger than the critical nucleus size (see the next section). In the case of the B33 nucleation under moderate undercoolings, the simulation cell contained ~1 million atoms which is around 18 times larger than the critical nucleus size. 

\section{Critical nucleus size and solid-liquid interface free energy}

\begin{figure}[t]
\includegraphics[width=0.48\textwidth]{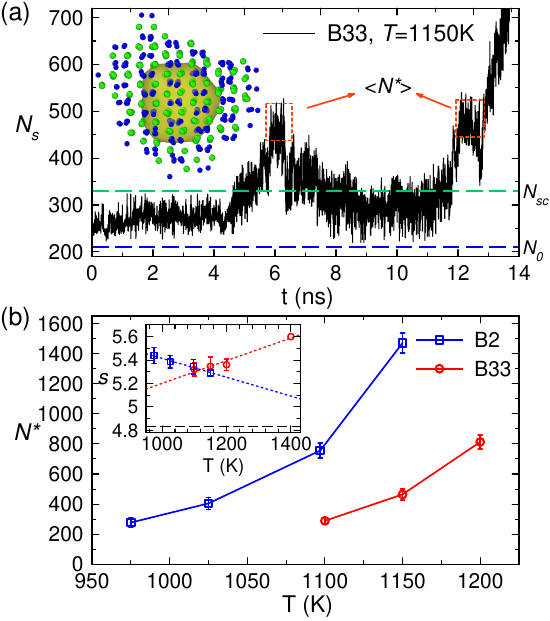}
\caption{\label{fig:fig3} (a) Nucleus size as a function of time in the PEM-MD simulation for NiZr-B33 at 1150 K. The red boxes indicate the plateau regions of the critical nucleus. The insert shows a snapshot of the critical nucleus from the first plateau; the Ni and Zr atoms are shown by the blue and green dots, respectively. The central yellow region indicates the initial embryo. (b) The critical nucleus sizes and shape factors (insert) for both NiZr-B2 and NiZr-B33 as functions of temperature. The error bars were obtained by measuring the multiple critical nuclei collected from PEM simulations. The data at 1400K are obtained from the seeding method. The dot lines indicate the linear fitting of the data. The black dashed line shows the perfect sphere shape factor.}
\end{figure} 

Figure~\ref{fig:fig3}(a) shows a typical result of the PEM-MD simulation. In this case, the initial embryo size $N_0$ is 210, and the threshold for removing the springs $N_{sc}$ is 330. The plateau indicates the appearance of the critical nucleus. The critical nucleus size $N^*$ can be directly measured by averaging the size at the plateau \cite{R27}. The temperature dependences of critical nucleus size for both the B2 and B33 phases are shown in Fig.~\ref{fig:fig3}(b). The shape factors were computed as $s=\frac{A}{V^{2/3}}$ \cite{R28}, where $A$ is the surface area and $V$ is the volume of the nucleus polyhedron constructed by the geometric surface reconstruction method \cite{R46} integrated in the OVITO software package \cite{R47}. The examination of Fig.~\ref{fig:fig3}(b) shows that the shape factors of both B2 and B33 critical nuclei deviate from the sphere even though the initial embryo was spherical (see insert in Fig.~\ref{fig:fig3}(a)). Interestingly, the shape factors of B2 and B33 exhibit different temperature dependences. The shape factor of the B2 phase decreases with increasing temperature indicating that the SLI becomes more anisotropic with increasing supercooling. This is similar to the previously measured shape factors for Al and Ni \cite{R28}. The variation of the B33 shape factor in the temperature range studied using the PEM is within the statistical uncertainty of the data. Therefore, based on these data we cannot make any conclusion about the temperature dependence of the anisotropy of the B33 SLI.
 
\begin{figure}
\includegraphics[width=0.48\textwidth]{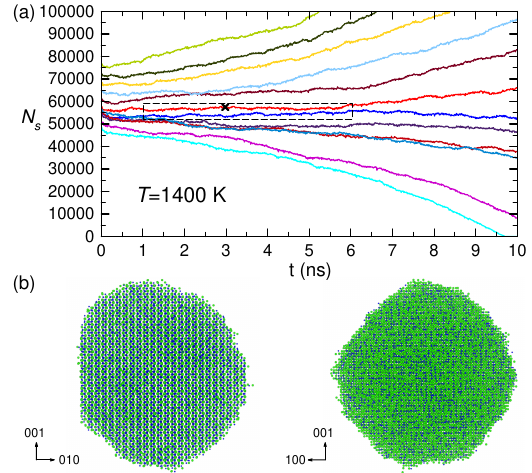}
\caption{\label{fig:fig4} Size evolutions for 12 nuclei with different initial sizes. The ones in the dashed box were used to estimate the critical nucleus size. (b) The projections of a critical nucleus from two orientations. The time when the nucleus was extracted are cross-marked in panel (a).}
\end{figure} 

To further evaluate the SLI free energy and the shape factor of the B33 phase, we approached a very small undercooling at 1400 K (5\% undercooling, ~60 K below the B33 melting temperature). The critical nucleus size of the B33 phase cannot be obtained under these conditions even from the PEM because as will be shown below the critical nucleus consists of ~50,000 atoms and the growth rate is very low. On the other hand, it is not obvious that the conventional seeding technique where a nucleus is embedded into the liquid and the critical size is determined by the fact whether the nucleus grows or disappears, is applicable to the phases with very large anisotropy of the SLI properties as was observed in \cite{R34}. Therefore, in the present study, we first let the nucleus obtained by the PEM grow to significantly large sizes at a lower temperature, $T=1150 K$. Then we increased the temperature to the target value ($T=1400 K$) and registered if the nuclei of different sizes grew or melted. Figure~\ref{fig:fig4}(a) shows that when the nucleus is too large or too small, it indeed grows or melts, respectively. Only when the size is near the critical size the nucleus neither grows nor melts for a quite long period. While this technique (which is in fact based on the assumption that the anisotropy of the SLI properties does not change below the target temperature) does not provide the same level of accuracy as the PEM, it can still be used to estimate the critical nucleus size. The critical nucleus size was determined as $N^*=55,500 \pm 1,400$ according to the two trajectories marked in Fig.~\ref{fig:fig4}(a). One of the obtained critical nuclei is shown in Fig.~\ref{fig:fig4}(b). As can be seen in this figure, the critical nucleus shape exhibits a quite large anisotropy. The SLI in the [010] direction is very flat which is consistent with the data obtained from the CFM at the melting point \cite{R34}. 

The increase in the SLI anisotropy can also be seen from the comparison of the shape factor values obtained from the seeding simulation at T=1400 K and from the PEM simulations at lower temperatures (see the insert in Fig.~\ref{fig:fig3}(b)). Although we have to note that the high temperature value was obtained based on just one simulation, the large critical nucleus size (~55,500 atoms) makes this value rather reliable. This increase in the shape factor with increasing temperature is opposite to the temperature dependences of the shape factors which we observed for pure Al and Ni in \cite{R28} and for the B2 phase in the present study, but it is consistent with the fact one interface was immobile in the MD simulations of the flat interface migration nearby the melting temperature while others were mobile but all interfaces migrated with about the same velocity at large supercoolings \cite{R34}. 

\begin{figure} 
\includegraphics[width=0.48\textwidth]{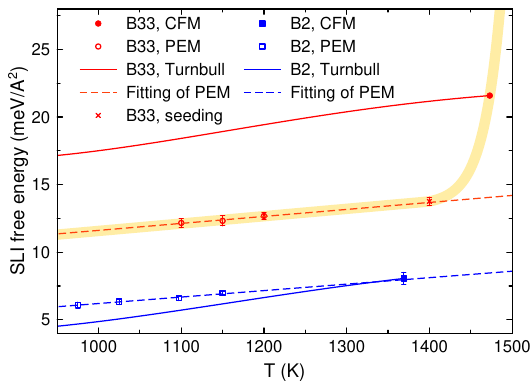} 
\caption{\label{fig:fig5} The SLI free energy as a function of temperature. The dashed lines are the linear fitting and extrapolations of the PEM data ($\gamma_{B2}=1.413+0.004784T (meV/ \AA^2)$ and $\gamma_{B33}=6.438+0.005173T (meV/\AA ^2)$). The solid lines are estimations by the Turnbull relation \cite{R3}. The yellow curve is just a guideline for the temperature dependence of B33 SLI free energy.}
\end{figure} 

Once both the critical nucleus size and the shape factor are determined, the SLI free energy can be calculated using Eq. (5). The results from the current PEM and seeding simulations as well as the previous CFM results \cite{R34} are summarized in Fig.~\ref{fig:fig5}. The SLI free energy obtained from current PEM simulations shows almost linear dependence on the temperature for both B2 and B33 phases. For the B2 phase, the linear extrapolation to the melting temperature agrees well with the CFM value. However, the linear extrapolation for the B33 phase largely deviates from the CFM value, while it agrees well with the value obtained from the seeding method. Figure~\ref{fig:fig5} also show the temperature dependence of the SLI free energy estimated by the Turnbull relation \cite{R33} using Eqn. (1). Clearly the predictions obtained from this relation significantly deviates from the data obtained from the current PEM simulations for both the B2 and B33 phases. Moreover, the Turnbull relation overestimates the SLI free energy for the B33 phase and underestimates it for the B2 phase.

\section{Nucleation rate} 
\begin{figure} [hbt!]
\includegraphics[width=0.47\textwidth]{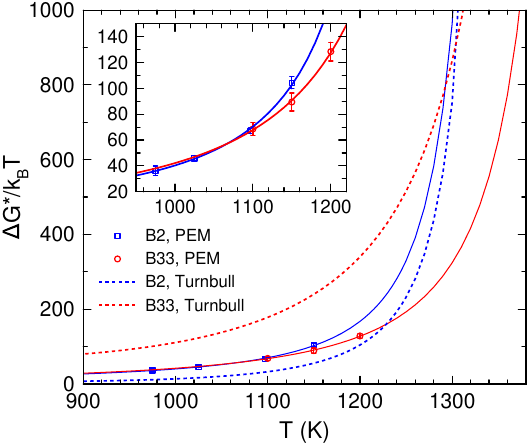}
\caption{\label{fig:fig6} The nucleation barriers as functions of temperature. The solid line are the extrapolation of PEM data with Eqn. (2). The insert zooms in the temperature regime where PEM was performed.}
\end{figure}

Based on the steady-state model \cite{R1}, the nucleation rate, $J$, can be expressed as 
\begin{equation}
J=\rho_L f^{+} \sqrt{\frac{\vert \Delta \mu \vert}{6 \pi k_{B} T N^{*}}} \exp(-\frac{\Delta G^{*}}{k_{B} T})
\label{eos},
\end{equation} 
where $k_B$ is the Boltzmann constant, $f^+$ is the attachment rate of a single atom to the critical nucleus and $\rho_L$ is the liquid density. Using the temperature dependence of the interfacial free energy, the nucleation barrier $\Delta G^*$  can be obtained from Eq. (3) for the same temperature range. Figure~\ref{fig:fig6} shows $\Delta G^*$ as function of temperature for both the B2 and B33 phases. The two curves cross near 1070 K and in fact they are statistically indistinguishable below this temperature. For comparison, we also include the free energy barriers computed by using the SLI free energy obtained from the Turnbull relation \cite{R33}. Obviously these data significantly deviate from the current PEM-MD results.

\begin{figure}
\includegraphics[width=0.48\textwidth]{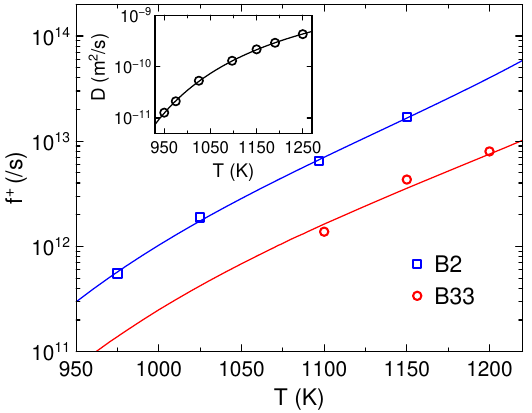}
\caption{\label{fig:fig7} The attachment rate obtained from Eq. (6). The squares and circles represent the data measured in the iso-configurational MD simulations \cite{R27}. By fitting the PEM data, $\lambda$ is determined as 1.43 \AA and 2.11 \AA for the B2 and B33 phases, respectively. The insert shows the VFT fitting of the diffusion coefficient for Zr in Ni$_{\text{50}}$Zr$_{\text{50}}$ liquid measured by MD simulations.}
\end{figure}

The attachment rate, $f^+$, was measured in the iso-configurational MD simulations as the effective diffusion constant for the change of the critical nucleus size \cite{R27,R48,R49}. The obtained MD results are shown in Fig.~\ref{fig:fig7} for both the B2 and B33 phases. To fit the data with the temperature dependence, we consider the classical kinetic model of atom attachment \cite{R1} where $f^+$ is proportional to the liquid diffusivity $D$ and the nucleus surface area. With the shape factor correction, the expression for $f^+$ can be written as
\begin{equation}
f^{+}=s^{*}N^{*2/3}\frac{6D}{\lambda^2 }
\label{eos}.
\end{equation} 
where $\lambda$ is the atomic jump distance during the attachment which can be determined based on the measured $f^+$. For the undercooled NiZr liquids considered here, the temperature dependence of the bulk diffusion coefficient can be well fit to the Vogel-Fulcher-Tammann (VFT) relationship \cite{R50}, which are shown in Fig.~\ref{fig:fig7} . Here we used the diffusivity of Zr atoms because it is slower than that of Ni atoms, and thus is the limiting factor of the attachment. With all the parameter in Eqn. (7) available, the attachment rate is extrapolated to a wide temperature range. The attachment rate on the B2 nucleus is typically 4-5 times faster than the one for the B33 nucleus at the same temperature.

Finally, once all the temperature dependences of the parameters in Eq. (6) were determined, the nucleation rate was computed for both the B2 and B33 phases (see Fig.~\ref{fig:fig8}). The two curves cross at 1070 K, where the nucleation rate of the B2 phase becomes larger than that for the B33 phase. Figure~\ref{fig:fig8}  also shows that the nucleation rate computed using $\gamma$ obtained from the CFM value and the Turnbull relation dramatically deviates from the PEM data.

\section{Discussion}
\begin{figure}
\includegraphics[width=0.48\textwidth]{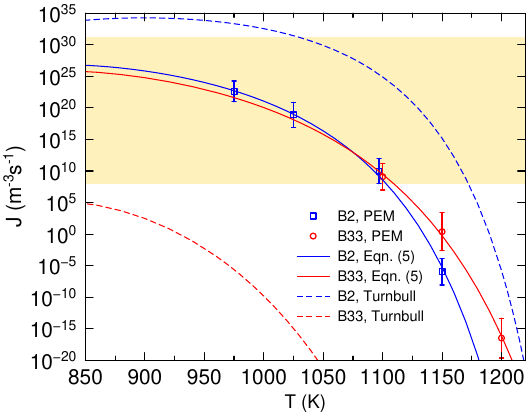}
\caption{\label{fig:fig8} The nucleation rate as a function of temperature. The yellow shadow indicates the region of homogeneous nucleation. The lower limit is estimated by the typical experimental conditions \cite{R33,R51}, while the upper limit is estimated for the brute-force MD simulations.}
\end{figure}

In the present study, we used the persistent embryo method and the classical nucleation theory to determine the nucleation rates for the B2 and B33 phases in the Ni$_{\text{50}}$Zr$_{\text{50}}$ alloy. The study was motivated by two questions: i) if it is really important to take into account the actual temperature dependence of the SLI free energy or using the CFM values obtained at the melting temperature can give a reliable prediction and ii) if the temperature dependence of the SLI free energy is always approximately linear as was obtained in \cite{R28,R29,R30} for rather simple crystal phases. These questions are especially important for understanding the nucleation in the Ni$_{\text{50}}$Zr$_{\text{50}}$ alloy. Until the levitation study performed at very large undercooling in \cite{R33} it was believed that this alloy should solidify directly into the B33 phase. While the levitation study demonstrated that it is the B2 phase which nucleates first at large undercoolings, it obviously cannot be the case in the temperature interval between the B2 and B33 melting temperatures. However, the calculations performed using the CFM data and the Turnbull relation suggest that the B33 phase can never homogeneously nucleate from the liquid. Indeed, with the experimental sample size of 2.0 mm, the nucleation rate which can be accessed in a few seconds must be higher than $10^8 m^{-3}s^{-1}$ (see the estimations made in \cite{R51}). This threshold value (the lower bound of the yellow range shown in Fig.~\ref{fig:fig8}) is several orders of magnitude higher than the values predicted by using the CFM data and the Turnbull relation (see Fig.~\ref{fig:fig8}). On contrary, the calculations performed used the PEM show that the B33 can homogeneously nucleate below $T=1100 K$. 

In reality it is very difficult to conclude which technique provides more reliable predictions by comparison with the experiment. In the levitation experiment reported in \cite{R33} the B2 nucleation happened at ~1350 K after which the sample heated up to ~1550 K which is probably the B2 melting temperature. However, the semi-empirical potential used in the present study provides that the B2 melting temperature is 1369 K (the authors of \cite{R34} did not have any data on the B2 melting temperature and simply made it lower than the B33 melting temperature; the target value for the B33 melting temperature was 1533 K in \cite{R34}). Thus, it is reasonable that our prediction for the beginning of the B2 homogeneous nucleation is ~250 K below the temperature where it was observed in the levitation experiment and the disagreement can be simply attributed to the inaccuracy of the employed semi-empirical potential. Our present calculations predict that there should be close competition between the nucleation of the B2 and B33 phases and indeed the B33 phase shows up in the experiment just a few seconds after the B2 phase nucleates. However, it is still unclear how the B33 phase nucleates in the experiment. If it nucleates from the liquid this would be indeed the case of a close competition because it happens when the B2 phase is still solidifying. If on the other hand, the B33 phase nucleates from the B2 phase the experiment does not provide any information about how close the competition between the B2 and B33 phase nucleation from the liquid is. Thus the only validation of our calculations from the experiment is that the B33 phase cannot nucleate before the B2 phase even at high temperatures where the B33 nucleation rate is higher than the B2 nucleation rate, because the B33 nucleation rate at these temperatures is too low. 

Another way to validate the accuracy of the predictions for the nucleation rate is to compare them with the results of the brute-force MD simulations at deep undercoolings \cite{R27,R28}. In this case an MD simulation is simply performed until the nucleation is observed and the inaccuracy of the employed semi-empirical potential does not play any role as long as the same semi-empirical potential is used in all calculations. Therefore, in order to compare which approach provides a better prediction we performed a brute-force MD simulation at $T=850 K$ using a simulation cell containing 31,250 atoms. Due to the stochastic nature of the nucleation, 10 independent runs were performed. Unfortunately, in spite of the fact that the duration of the simulations was 200 $ns$, no nucleation was observed. Therefore, the nucleation rate at $T=850 K$ with the current the semi-empirical potential must be lower than $10^{31} m^{-3}s^{-1}$ (the upper bound of the yellow range shown in Fig.~\ref{fig:fig8}). The calculations made using the PEM show that the nucleation rate never reach this value and therefore, indeed predict that no nucleation should be observed in the MD simulation. On contrary, the calculation based on the CFM value and the Turnbull relation leads to the prediction that the nucleation should be observed in the MD simulation below ~1025 K. This vividly demonstrates that an accurate determination of SLI free energy is absolutely necessary to get an accurate prediction for the nucleation rate from the CNT.

Figure~\ref{fig:fig5} shows that the temperature dependence of the SLI free energy of the B2 phase is linear just like in the cases of pure Al and Ni considered in  \cite{R28}. However, in the case of more complex B33 phase the temperature dependence is not that trivial. In fact, it remains unclear how the B33 SLI free energy depends on the temperature in the range from 1400 K to the B33 melting temperature, 1473 K. Obviously it becomes very non-linear which is intuitively consistent with the observation from the flat SLI migration simulations where an anomalous non-moving B33 orientation was found at the temperature range from 1450 K to 1490 K \cite{R34}. Unfortunately, this temperature region could not be probed using the PEM because of the too large computational cost.

Finally, we note that the SLI free energies obtained in the present study are averaged over all crystal orientations and the anisotropy of the SLI free energy was neglected. The seeding simulations at $T=1400 K$ revealed a strongly anisotropic shape of the B33 nucleus even though the critical nucleus was very large. Yet the SLI free energy estimated from the seeding simulations still agrees with the linear extrapolation from the PEM results (see Fig.~\ref{fig:fig5}). 

\section{Conclusion}
In summary, using the CNT and PEM, we determined the SLI free energies and nucleation rates for the B2 and B33 phases in the Ni$_{\text{50}}$Zr$_{\text{50}}$ alloy. The competition between the B2 and B33 phase nucleation was quantified and the results are consistent with the experimental observations. In the case of the B2 phase, the linear temperature dependence of the SLI free energy extrapolated to the melting temperature leads to the same value as the one obtained from the CFM. In the case of the B33 phases, the SLI free energy is also a linear function of temperature at large supercoolings but the extrapolation to the melting temperature leads to a value which is considerably different from the CFM value. This is consistent with the large anisotropy of the SLI properties of the B33 phase near the melting temperature. 

\section{Acknowledgments}
Work at Ames Laboratory was supported by the U.S. Department of Energy (DOE), Office of Science, Basic Energy Sciences, Materials Science and Engineering Division, under Contract No. DE-AC02-07CH11358, including a grant of computer time at the National Energy Research Supercomputing Center (NERSC) in Berkeley, CA. The Laboratory Directed Research and Development (LDRD) program of Ames Laboratory supported the use of GPU computing.

\end{document}